\documentclass[aps,pre,twocolumn,amsmath,showpacs]{revtex4}
\usepackage{graphicx}
\begin{document}

\title{Critical behavior and Griffiths effects in the disordered contact process}

\author{Thomas Vojta}
\author{Mark Dickison}
\affiliation{Department of Physics, University of Missouri-Rolla, Rolla, MO 65409}

\date{\today}

\begin{abstract}
We study the nonequilibrium phase transition in the one-dimensional contact process with
quenched spatial disorder by means of large-scale Monte-Carlo simulations for times up to
$10^9$ and system sizes up to $10^7$ sites. In agreement with recent predictions of an
infinite-randomness fixed point, our simulations demonstrate activated (exponential)
dynamical scaling at the critical point. The critical behavior turns out to be universal,
even for weak disorder. However, the approach to this asymptotic behavior is extremely
slow, with crossover times of the order of $10^4$ or larger. In the Griffiths region
between the clean and the dirty critical points, we find power-law dynamical behavior
with continuously varying exponents. We discuss the generality of our findings and relate
them to a broader theory of rare region effects at phase transitions with quenched
disorder.

\end{abstract}



\maketitle

\section{Introduction}
\label{sec:intro}

The nonequilibrium behavior of many-particle systems has attracted considerable attention
in recent years. Of particular interest are continuous phase transitions between
different nonequilibrium states. These transitions are characterized by large scale
fluctuations and collective behavior over large distances and times very similar to the
behavior at equilibrium critical points. Examples of such nonequilibrium transitions can
be found in population dynamics and epidemics, chemical reactions, growing surfaces, and
in granular flow and traffic jams (for recent reviews see, e.g., Refs.\
\cite{SchmittmannZia95,Marro_book,Dickman,chopard_book,Hinrichsen00,Odor,tauber_rev})

A prominent class of nonequilibrium phase transitions separates active fluctuating states
from inactive, absorbing states where fluctuations cease entirely. Recently, much effort
has been devoted to classifying possible universality classes of these absorbing state
phase transitions \cite{Hinrichsen00,Odor}. The generic universality class is directed
percolation (DP) \cite{dp}. According to a conjecture by Janssen and Grassberger
\cite{conjecture}, all absorbing state transitions with a scalar order parameter,
short-range interactions, and no extra symmetries or conservation laws belong to this
class. Examples include the transitions in the contact process \cite{contact}, catalytic
reactions \cite{ziff}, interface growth \cite{tang}, or turbulence \cite{turb}. In the
presence of conservation laws or additional symmetries, other universality classes can
occur, e.g., the parity conserving class \cite{Takayasu92,Zhong95,Tauber96} or the
$Z_2$-symmetric directed percolation (DP2) class
\cite{KimPark94,Menyhard94,Hinrichsen97}.

In realistic systems, one can expect impurities and defects, i.e., quenched spatial
disorder, to play an important role. Indeed, it has been suggested that disorder may be
one of the reasons for the surprising rarity of experimental realizations of the
ubiquitous directed percolation universality class \cite{hinrichsen_exp}. The only
verification so far seems to be found in the spatio-temporal intermittency in
ferrofluidic spikes \cite{spikes}.

The investigation of disorder effects on the DP transition has a long history, but a
coherent picture has been slow to emerge. According to the Harris criterion
\cite{harris,Noest86}, a clean critical point is stable against weak disorder if the
spatial correlation length critical exponent $\nu_\perp$ fulfills the inequality
\begin{equation}
d\nu_\perp > 2,
\end{equation}
where $d$ is the spatial dimensionality. The DP universality class violates the Harris
criterion in all dimensions $d<4$, because the exponent values are $\nu_\perp \approx
1.097$ (1D), 0.73 (2D), and 0.58 (3D) \cite{Hinrichsen00}. A field-theoretic
renormalization group study \cite{janssen97} confirmed the instability of the DP critical
fixed point. Moreover, no new critical fixed point was found. Instead the renormalization
group displays runaway flow towards large disorder, indicating unconventional behavior.
Early Monte-Carlo simulations \cite{Noest86} showed significant changes in the critical
exponents while later studies \cite{moreira} of the two-dimensional contact process with
dilution found logarithmically slow dynamics in violation of power-law scaling. In
addition, rare region effects similar to Griffiths singularities \cite{Griffiths} were
found to lead to slow dynamics in a whole parameter region in the vicinity of the phase
transition \cite{moreira,Noest88,bramson,webman,cafiero}.

Recently, an important step towards understanding spatial disorder effects on the DP
transition has been made by Hooyberghs et al.\ \cite{hooyberghs}. These authors used the
Hamiltonian formalism \cite{alcaraz} to map the one-dimensional disordered contact
process onto a random quantum spin chain. Applying a version of the Ma-Dasgupta-Hu
strong-disorder renormalization group \cite{SDRG}, they showed that the transition is
controlled by an infinite-randomness critical point, at least for sufficiently strong
disorder. This type of fixed point leads to activated (exponential) rather than power-law
dynamical scaling. For weaker disorder, Hooyberghs et al.\ \cite{hooyberghs} used
computer simulations and predicted non-universal continuously varying exponents, with
either power-law or exponential dynamical correlations.

In this paper, we present the results of large-scale Monte-Carlo simulations of the
one-dimensional contact process with quenched spatial disorder. Using large systems of up
to $10^7$ sites and very long times (up to $10^9$) we show that the critical behavior at
the nonequilibrium phase transition is indeed described by an infinite-randomness fixed
point with activated scaling. Moreover, we provide evidence that this behavior is
universal, i.e., it occurs even in the weak-disorder case. However, the approach to this
universal asymptotic behavior is extremely slow, with crossover times of the order of
$10^4$ or larger which may explain why nonuniversal (effective) exponents have been seen
in previous work. We also study the Griffiths region between the clean and the dirty
critical points. Here, we find power-law dynamical behavior with continuously varying
exponents, in agreement with theoretical predictions.

This paper is organized as follows. In section \ref{sec:theory}, we introduce the model.
We then contrast power-law scaling as found at conventional critical points with
activated scaling arising from infinite-randomness critical points. We also summarize the
predictions for the Griffiths region. In section \ref{sec:mc}, we present our simulation
method and the numerical results together with a comparison to theory. We conclude in
section \ref{sec:conclusions} by discussing the generality of our findings and their
relation to a broader theory of rare region effects at phase transitions with quenched
disorder.

\section{Theory}
\label{sec:theory}
\subsection{Contact process with quenched spatial disorder}

We start from the clean contact process \cite{contact}, a prototypical system in the DP
universality class. It can be interpreted, e.g., as a model for the spreading of a
disease. The contact process is defined on a $d$-dimensional hypercubic lattice. Each
lattice site $\mathbf r$ can be active (occupied by a particle) or inactive (empty). In
the course of the time evolution, active sites can infect their neighbors, or they can
spontaneously become inactive. Specifically, the dynamics is given by a continuous-time
Markov process during which particles are created at empty sites at a rate $\lambda n/
(2d)$ where $n$ is the number of active nearest neighbor sites. Particles are annihilated
at rate $\mu$ (which is often set to unity without loss of generality). The ratio of the
two rates controls the behavior of the system.

For small birth rate $\lambda$, annihilation dominates, and the absorbing state without
any particles is the only steady state (inactive phase). For large birth rate $\lambda$,
there is a steady state with finite particle density (active phase).  The two phases are
separated by a nonequilibrium phase transition in the DP universality class at
$\lambda=\lambda_c^0$. The central quantity in the contact process is the average density
of active sites at time $t$
\begin{equation}
\rho(t) = \frac 1 {L^d} \sum_{\mathbf{r}} \langle  n_\mathbf{r}(t) \rangle
\end{equation}
where $n_\mathbf{r}(t)$ is the particle number at site $\mathbf{r}$ and time $t$, $L$ is
the linear system size, and $\langle \ldots \rangle$ denotes the average over all
realizations of the Markov process. The longtime limit of this density (i.e., the steady
state density)
\begin{equation}
\rho_{\rm stat} = \lim_{t\to\infty} \rho(t)
\end{equation}
is the order parameter of the nonequilibrium phase transition.

Quenched spatial disorder can be introduced by making the birth rate $\lambda$ a random
function of the lattice site $\mathbf{r}$. We assume the disorder to be spatially
uncorrelated; and we use a binary probability distribution
\begin{equation}
P[\lambda({\mathbf r})] = (1-p)\, \delta[\lambda({\mathbf r})-\lambda] + p\,
\delta[\lambda({\mathbf r}) - c\lambda] \label{eq:impdist}
\end{equation}
where $p$ and $c$ are constants between 0 and 1. This distribution allows us to
independently vary spatial density $p$ of the impurities and their relative strength $c$.
The impurities locally \emph{reduce} the birth rate, therefore, the nonequilibrium
transition will occur at a value $\lambda_c$ that is larger than the clean critical birth
rate $\lambda_c^0$.

\subsection{Conventional power-law scaling}

In this subsection we summarize the phenomenological scaling theory of an absorbing state
phase transition that is controlled by a conventional fixed point with power-law scaling
(see, e.g., Ref.\ \cite{Hinrichsen00}). The clean contact process falls into this class.

In the active phase and close to critical point $\lambda_c$, the order parameter
$\rho_{\rm stat}$ varies according to the power law
\begin{equation}
\rho_{\rm stat} \sim (\lambda-\lambda_c)^\beta \sim \Delta^\beta
\end{equation}
where $\Delta=(\lambda-\lambda_c)/\lambda_c$ is the dimensionless distance from the
critical point, and $\beta$ is the critical exponent of the particle density. In addition
to the average density, we also need to characterize the length and time scales of the
density fluctuations. Close to the transition, the correlation length $\xi_\perp$
diverges as
\begin{equation}
\xi_\perp \sim |\Delta|^{-\nu_{\perp}}~.
\end{equation}
The correlation time $\xi_\parallel$ behaves like a power of the correlation length,
\begin{equation}
\xi_\parallel \sim \xi_\perp^z, \label{eq:powerlawscaling}
\end{equation}
i.e., the dynamical scaling is of power-law form. Consequently the scaling form of the
density as a function of $\Delta$, the time $t$ and the linear system size $L$ reads
\begin{equation}
\rho(\Delta,t,L) = b^{\beta/\nu_\perp} \rho(\Delta b^{-1/\nu_\perp},t b^z, L b)~.
\label{eq:rho}
\end{equation}
Here, $b$ is an arbitrary dimensionless scaling factor.

Two important quantities arise from initial conditions consisting of a single active site
in an otherwise empty lattice. The survival probability $P_s$ describes the probability
that an active cluster survives when starting from such a single-site seed. For directed
percolation, the survival probability scales exactly like the density \cite{betaprime},
\begin{equation}
P_s(\Delta,t,L) = b^{\beta/\nu_\perp} P_s(\Delta b^{-1/\nu_\perp},t b^z, L b)~.
\label{eq:Ps}
\end{equation}
Thus, for directed percolation, the three critical exponents $\beta$, $\nu_\perp$ and $z$
completely characterize the critical point.

 The pair connectedness function $C(\mathbf{r'},t',\mathbf{r},t)=\langle
n_{\mathbf{r'}}(t') \, n_{\mathbf{r}}(t) \rangle$ describes the probability that site
$\mathbf{r}'$ is active at time $t'$ when starting from an initial condition with a
single active site at $\mathbf{r}$ and time $t$. For a clean system, the pair
connectedness is translationally invariant in space and time. Thus, it only depends on
two arguments $C(\mathbf{r},t',\mathbf{r},t)=C(\mathbf{r'}-\mathbf{r},t-t')$. Because $C$
involves a product of two densities, its scale dimension is $2\beta/\nu_\perp$, and the
full scaling form reads \cite{hyperscaling}
\begin{equation}
C(\Delta,\mathbf{r},t,L) = b^{2\beta/\nu_\perp} C(\Delta b^{-1/\nu_\perp}, \mathbf{r}b, t
b^z, L b)~.
\end{equation}
The total number of particles $N$ when starting from a single seed site can be obtained
by integrating the pair connectedness $C$ over all space. This leads to the scaling form
\begin{equation}
N(\Delta,t,L) = b^{2\beta/\nu_\perp - d} N(\Delta b^{-1/\nu_\perp},t b^z, L b)~.
\label{eq:N}
\end{equation}

At the critical point, $\Delta=0$, and in the thermodynamic limit, $L\to\infty$, the
above scaling relations lead to the following predictions for the time dependencies of
observables: The density and the survival probability asymptotically decay like
\begin{equation}
\rho(t) \sim t^{-\delta}, \qquad P_s(t) \sim t^{-\delta}
\end{equation}
with $\delta=\beta/(\nu_\perp z)$. In contrast, the number of particles in a cluster
starting from a single seed site increases like
\begin{equation}
N(t) \sim t^\Theta
\end{equation}
where $\Theta=d/z - 2\beta/(\nu_\perp z)$ is the so-called critical initial slip
exponent.

Highly precise estimates of the critical exponents for clean one-dimensional directed
percolation have been obtained by series expansions \cite{Jensen99}: $\beta=0.276486$,
$\nu_\perp=1.096854$, $z=1.580745$, $\delta=0.159464$, and $\Theta=0.313686$.

\subsection{Activated scaling}
\label{subsec:activated}

In this subsection we summarize the scaling theory for an infinite-randomness fixed point
with activated scaling, as has been predicted to occur in absorbing state transitions
with quenched disorder \cite{hooyberghs}. It is similar to the scaling theory for the
quantum phase transition in the random transverse field Ising model \cite{dsf9295}.

At an infinite-randomness fixed point, the dynamics is extremely slow. The power-law
scaling (\ref{eq:powerlawscaling}) gets replaced by activated dynamical scaling
\begin{equation}
\ln(\xi_\parallel) \sim \xi_\perp^\psi, \label{eq:activatedscaling}
\end{equation}
characterized by a new exponent $\psi$. This exponential relation between time and length
scales implies that the dynamical exponent $z$ is formally infinite. In contrast, the
static scaling behavior remains of power law type.

Moreover, at an infinite-randomness fixed point the probability distributions of
observables become extremely broad, so that averages are dominated by rare events such as
rare spatial regions with large infection rate. In such a situation, averages and typical
values of a quantity do not necessarily agree. Nonetheless, the scaling form of the
\emph{average} density at an infinite-randomness critical point is obtained by simply
replacing the power-law scaling combination $t b^z$ by the activated combination $\ln(t)
b^\psi$ in the argument of the scaling function:
\begin{equation}
\rho(\Delta,\ln(t),L) = b^{\beta/\nu_\perp} \rho(\Delta b^{-1/\nu_\perp},\ln(t) b^\psi, L
b)~. \label{eq:rho_activated}
\end{equation}
Analogously, the scaling forms of the average survival probability and the average number
of sites in a cluster starting from a single site are
\begin{eqnarray}
P_s(\Delta,\ln(t),L) &=& b^{\beta/\nu_\perp} P_s(\Delta b^{-1/\nu_\perp},\ln(t) b^\psi,L
b) \label{eq:Ps_activated}\\ N(\Delta,\ln(t),L) &=& b^{2\beta/\nu_\perp -d} N(\Delta
b^{-1/\nu_\perp},\ln(t) b^\psi,L b) ~~. \label{eq:N_activated}
\end{eqnarray}

These activated scaling forms lead to logarithmic time dependencies at the critical point
(in the thermodynamic limit). The average density and the survival probability
asymptotically decay like
\begin{equation}
\rho(t) \sim [\ln(t)]^{-\bar\delta}, \qquad P_s(t) \sim [\ln(t)]^{-\bar\delta}
\label{eq:logdecay}
\end{equation}
with $\bar\delta=\beta/(\nu_\perp \psi)$ while the average number of particles in a
cluster starting from a single seed site increases like
\begin{equation}
N(t) \sim [\ln(t)]^{\bar\Theta} \label{eq:clustersize}
\end{equation}
with $\bar\Theta=d/\psi-2\beta/(\nu_\perp \psi)$. These are the relations we are going to
test in this paper.

Within the strong disorder renormalization group approach of Ref.\ \cite{hooyberghs}, the
critical exponents of the disordered one-dimensional contact process can be calculated
exactly. Their numerical values are $\beta=0.38197$, $\nu_\perp=2$, $\psi=0.5$,
$\bar\delta=0.38197$, and $\bar\Theta=1.2360$.

\subsection{Griffiths region}
\label{subsec:Griffiths}

The inactive phase of our disordered contact process with the impurity  distribution
(\ref{eq:impdist}) can be divided into two regions. For birth rates below the clean
critical point, $\lambda<\lambda_c^0$, the behavior is conventional. The system
approaches the absorbing state exponentially fast in time. The decay time increases with
$\lambda$ and diverges as $|\lambda-\lambda_c^0|^{-z\nu_\perp}$ where $z$ and $\nu_\perp$
are the exponents of the clean critical point \cite{Bray88,Dickison05}.

The more interesting region is the so-called Griffiths region \cite{Noest86,Griffiths}
which occurs for birth rates between the clean and the dirty critical points,
$\lambda_c^0 < \lambda < \lambda_c$. The system is globally still in the inactive phase,
i.e., the system eventually decays into the absorbing state. However, in the
thermodynamic limit, one can find arbitrarily large spatial regions devoid of impurities.
For $\lambda_c^0 < \lambda < \lambda_c$, these so-called rare regions are locally in the
active phase. Because they are of finite size, they cannot support a non-zero steady
state density but their decay is very slow because it requires a rare, exceptionally
large density fluctuation.

The contribution of the rare regions to the time evolution of the density can be
estimated as follows \cite{Noest86,Noest88}. The probability $w$ for finding a rare
region of linear size $L_r$ devoid of impurities is (up to pre-exponential factors) given
by
\begin{equation}
 w(L_r) \sim \exp( -\tilde p L_r^d)
\end{equation}
where $\tilde p$ is a nonuniversal constant which for our binary disorder distribution is
given by $\tilde p = - \ln(1-p)$. The long-time decay of the density is dominated by
these rare regions. To exponential accuracy, the rare region contribution to the density
can be written as
\begin{equation}
\rho(t) \sim \int dL_r ~L_r^d ~w(L_r) \exp\left[-t/\tau(L_r)\right] \label{eq:rrevo}
\end{equation}
where $\tau(L_r)$ is the decay time of a rare region of size $L_r$. Let us first discuss
the behavior at the clean critical point, $\lambda_c^0$, i.e., at the boundary between
the conventional inactive phase and the Griffiths region. At this point, the decay time
of a single, impurity-free rare region of size $L_r$ scales as $\tau(L_r) \sim L_r^z$ as
follows from finite size scaling \cite{barber}. Here $z$ is the clean critical exponent.
Using the saddle point method to evaluate the integral (\ref{eq:rrevo}), we find the
leading long-time decay of the density to be given by a stretched exponential,
\begin{equation}
\ln \rho(t) \sim - \tilde{p}^{z/(d+z)}~ t^{d/(d+z)}~, \label{eq:stretched}
\end{equation}
rather than a simple exponential decay as for $\lambda<\lambda_c^0$.

Inside the Griffiths region, i.e., for $\lambda_c^0<\lambda<\lambda_c$, the decay time of
a single rare region depends exponentially on its volume,
\begin{equation}
\tau(L_r) \sim \exp(a L_r^d)
\end{equation}
because a coordinated fluctuation of the entire rare region is required to take it to the
absorbing state \cite{Noest86,Noest88,Schonmann85}. The nonuniversal prefactor $a$
vanishes at the clean critical point $\lambda_c^0$ and increases with $\lambda$. Close to
$\lambda_c^0$, it behaves as $a \sim \xi_\perp^{-d} \sim
(\lambda-\lambda_c^0)^{d\nu_\perp}$ with $\nu_\perp$ the clean critical exponent.
Repeating the saddle point analysis of the integral (\ref{eq:rrevo}) for this case, we
obtain a power-law decay of the density
\begin{equation}
\rho(t) \sim  t^{-\tilde p/a}  = t^{-d/z'} \label{eq:griffithspower}
\end{equation}
where $z'=da/\tilde p$ is a customarily used nonuniversal dynamical exponent in the
Griffiths region. Its behavior close to the \emph{dirty} critical point $\lambda_c$ can
be obtained within the strong disorder renormalization group method
\cite{hooyberghs,dsf9295}. When approaching the phase transition, $z'$ diverges as $z'
\sim  |\lambda-\lambda_c|^{-\psi\nu_\perp}$ where $\psi$ and $\nu_\perp$ are the
exponents of the dirty critical point.

\section{Monte-Carlo simulations}
\label{sec:mc}

\subsection{Method and overview}

We now turn to the main part of the paper, extensive Monte-Carlo simulations of the
one-dimensional contact process with quenched spatial disorder. There is a number of
different ways to actually implement the contact process on the computer (all equivalent
with respect to the universal behavior). We follow the widely used algorithm described,
e.g., by Dickman \cite{dickman99}. Runs start at time $t=0$ from some configuration of
occupied and empty sites. Each event consists of randomly selecting an occupied site
$\mathbf{r}$ from a list of all $N_p$ occupied sites, selecting a process: creation with
probability $\lambda(\mathbf{r})/[1+ \lambda(\mathbf{r})]$ or annihilation with
probability $1/[1+ \lambda(\mathbf{r})]$ and, for creation, selecting one of the
neighboring sites of $\mathbf{r}$. The creation succeeds, if this neighbor is empty. The
time increment associated with this event is $1/N_p$. Note that in this implementation of
the disordered contact process both the creation rate and the annihilation rate vary from
site to site in such a way that their sum is constant (and equal to one).

Using this algorithm, we have performed simulations for system sizes between $L=1000$ and
$L=10^7$. We have studied impurity concentrations $p=0.2, 0.3, 0.4, 0.5, 0.6$ and 0.7 as
well as relative impurity strengths of $c=0.2, 0.4, 0.6$ and 0.8. To explore the
extremely slow dynamics associated with the predicted infinite-randomness critical point,
we have simulated very long times up to $t=10^9$ which is, to the best of our knowledge,
at least three orders of magnitude in $t$ longer than previous simulations of the
disordered contact process. In all cases we have averaged over a large number of
different disorder realizations, details will be mentioned below for each specific set of
calculations.

\begin{figure}
\includegraphics[width=\columnwidth]{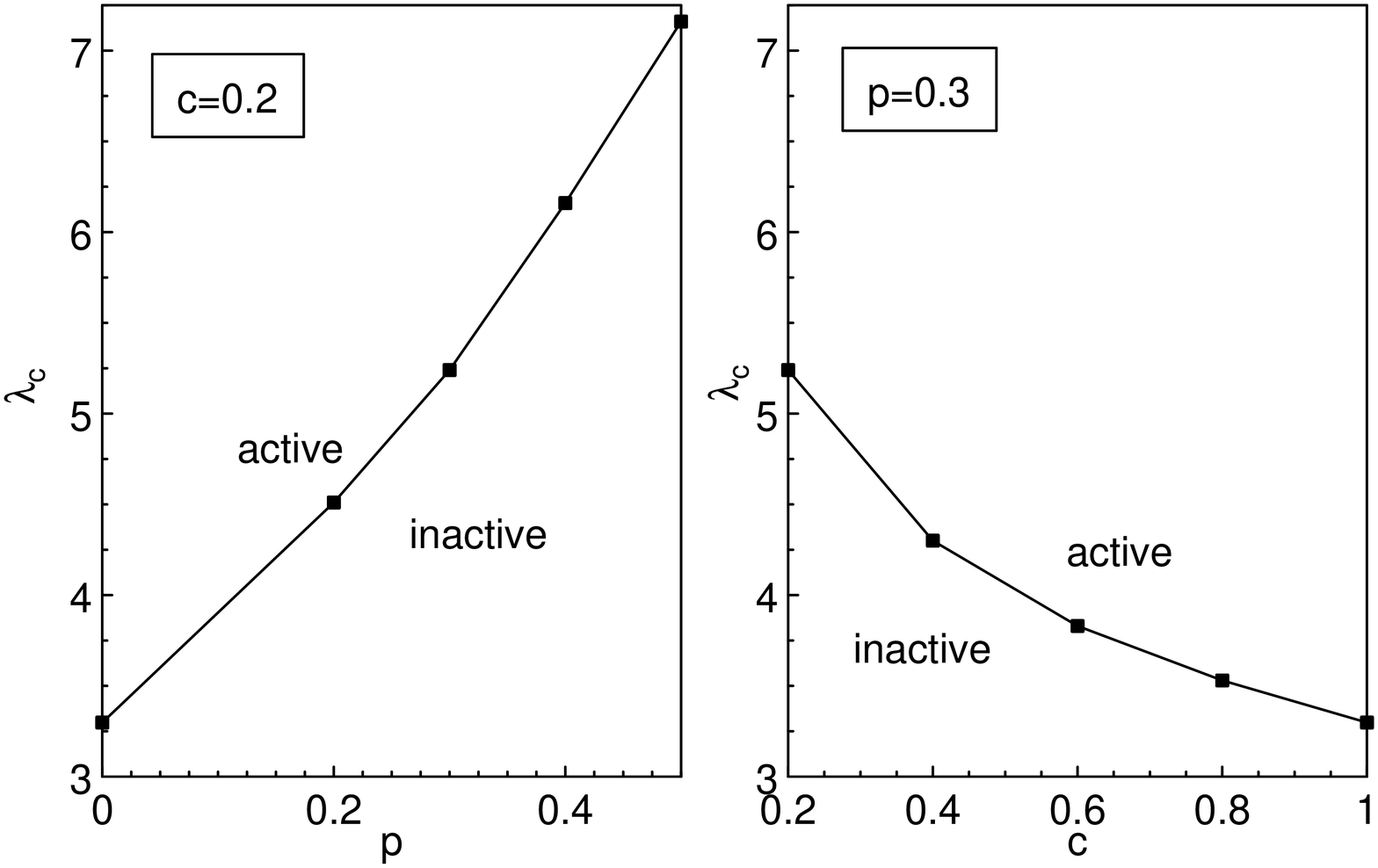}
\caption{Phase diagrams of the disordered contact process. Left: Birth rate $\lambda$ vs.
impurity concentration $p$ for fixed impurity strength $c=0.2$. Right: $\lambda$ vs. $c$
for fixed $p=0.3$.} \label{fig:pd}
\end{figure}

Figure \ref{fig:pd} gives an overview over the phase diagram resulting from our
simulations.
As expected, the critical birthrate $\lambda_c$ increases with increasing impurity
concentration $p$. It also increases with decreasing birth rate on the impurities, i.e.,
a decreasing relative strength $c$. For $p=0$ or $c=1$, we reproduce the well-known clean
critical birth rate $\lambda_c^0 \approx 3.298$ \cite{Jensen93}. In the following
subsections we discuss the behavior in the vicinity of the phase transition in more
detail.

\subsection{Time evolution starting from full lattice}

In this subsection we discuss simulations which follow the time evolution of the average
density starting from a full lattice. This means, at time $t=0$, all sites are active and
$\rho(0)=1$.

\begin{figure}
\includegraphics[width=\columnwidth]{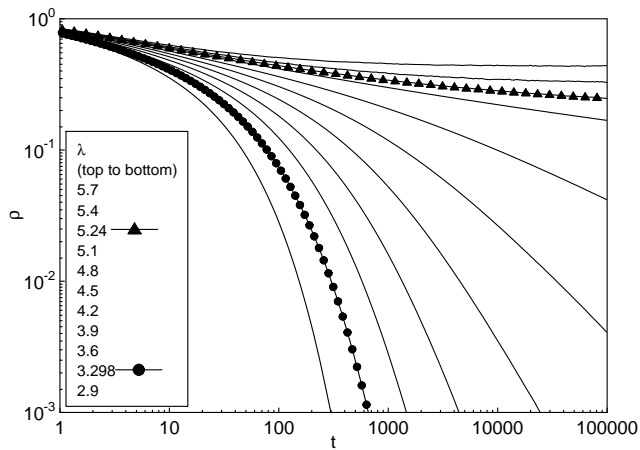}
\caption{Overview of the time evolution of the density for a system of $10^6$ sites with
$p=0.3$ and $c=0.2$. The clean critical point $\lambda_c^0\approx 3.298$ and the dirty
critical point $\lambda_c \approx 5.24$ are specially marked.} \label{fig:overviewp03c02}
\end{figure}

Figure \ref{fig:overviewp03c02} gives an overview of the time evolution of the density
for a system of $10^6$ sites with $p=0.3,c=0.2$, covering the $\lambda$ range from the
conventional inactive phase, $\lambda < \lambda_c^0$ all the way to the active phase,
$\lambda>\lambda_c$.
The data are averages over 480 runs, each with a different disorder realization. For
birth rates below and at the clean critical point $\lambda_c^0\approx 3.298$, the density
decay is very fast, clearly faster then a power law. Above $\lambda_c^0$, the decay
becomes slower and asymptotically seems to follow a power-law. For even larger birth
rates the decay seems to be slower than a power law while the largest birth rates give
rise to a nonzero steady state density, i.e., the system is in the active phase.

\subsubsection{Griffiths region}

\begin{figure}
\includegraphics[width=\columnwidth]{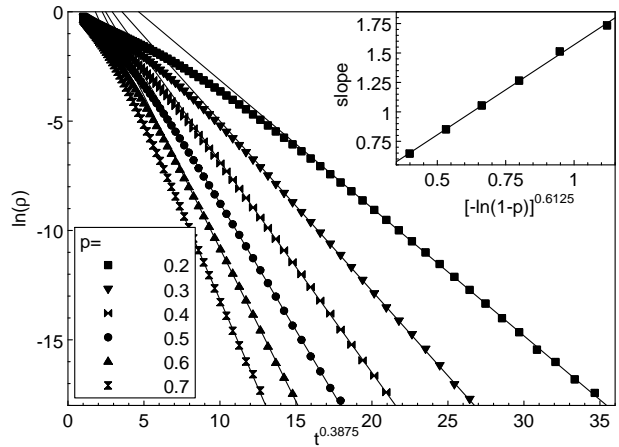}
\caption{Time evolution of the density at the clean critical point $\lambda_c^0=3.298$
for systems of $10^7$ sites with $c=0.2$ and several $p$.  The straight lines are fits to
the stretched exponential $\ln \rho(t) \sim -E t^{d/(d+z)}$ predicted in eq.\
(\ref{eq:stretched}) with $d=1$ and the clean $z=1.580$. Inset: Decay constant $E$ vs.
$\tilde{p}^{z/(d+z)}$. } \label{fig:evo_lc0}
\end{figure}

Let us investigate the different parameter regions in more detail, beginning with the
behavior at the clean critical point $\lambda_c^0$, i.e., at the boundary between the
Griffiths region and the conventional absorbing phase. According to eq.\
(\ref{eq:stretched}), the density should asymptotically decay like a stretched
exponential. To test this behavior, we plot the logarithm of the density as a function of
$t^{d/(d+z)}$ where $d=1$ and $z\approx 1.581$ is the dynamical exponent of the clean
one-dimensional contact process. Figure \ref{fig:evo_lc0} shows the resulting graphs for
system size $L=10^7$, $c=0.2$, and several impurity concentrations $p=0.2 \ldots 0.7$.
The data are averages over 960 runs, each with a different disorder realization. The
figure shows that the data follow a stretched exponential behavior $\ln \rho = -E
t^{0.3875}$ over more than four orders of magnitude in $\rho$, in good agreement with
eq.\ (\ref{eq:stretched}). The decay constant $E$, i.e., the slope of these curves,
increases with increasing impurity concentration $p$. The inset of figure
\ref{fig:evo_lc0} shows  the relation between $E$ and $\tilde p =-\ln(1-p)$. In good
approximation, the values follow the power law $E \sim \tilde p^{z/(d+z)}=\tilde
p^{0.6125}$ predicted in (\ref{eq:stretched}).

\begin{figure}
\includegraphics[width=\columnwidth]{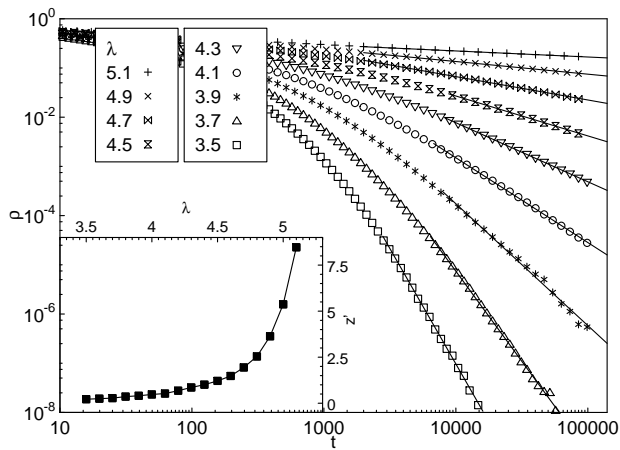}
\caption{Log-log plot of the density time evolution in the Griffiths region for systems
with $p=0.3, c=0.2$ and several birth rates $\lambda$. The system sizes are $10^7$ sites
for $\lambda=3.5, 3.7$ and $10^6$ sites for the other $\lambda$ values. The straight
lines are fits to the power law $\rho(t) \sim t^{-1/z'}$ predicted in eq.\
(\ref{eq:griffithspower}). Inset: Dynamical exponent $z'$ vs. birth rate $\lambda$.}
\label{fig:griffiths}
\end{figure}

We now turn to the behavior inside the Griffiths region, $\lambda_c^0 < \lambda <
\lambda_c$. Figure \ref{fig:griffiths} shows a double-logarithmic plot of the density
time evolution for birth rates $\lambda = 3.5 \ldots 5.1$ and $p=0.3, c=0.2$. The system
sizes are between $10^6$ and $10^7$ lattice sites, and we have averaged over 480 disorder
realizations.
For all birth rates $\lambda$ shown, the long-time decay of the density asymptotically
follows a power-law, as predicted in eq.\ (\ref{eq:griffithspower}), over several orders
of magnitude in $\rho$  (except for the largest $\lambda$ where we could observe the
power law only over a smaller range in $\rho$ because the decay is too slow).

The nonuniversal dynamical exponent $z'$ can be obtained by fitting the long-time
asymptotics of the curves in figure \ref{fig:griffiths} to eq.\
(\ref{eq:griffithspower}). The inset of figure \ref{fig:griffiths} shows $z'$ as a
function of the birth rate $\lambda$. As discussed in section \ref{subsec:Griffiths},
$z'$ increases with increasing $\lambda$ throughout the Griffiths region with an apparent
divergence around $\lambda \approx 5.2$. Unfortunately, our data did not allow us to make
quantitative comparisons with the predictions for the $\lambda$-dependence of $z'$
because we could not reliably determine $z'$ sufficiently close to either the clean
critical point or the dirty critical point.  Close to the clean critical point
$\lambda_c^0$, the crossover to the asymptotic power law occurs at very low densities,
thus the system size limits how close one can get to $\lambda_c^0$. Conversely, for
larger $\lambda$ close to the dirty critical point $\lambda_c$, the crossover to the
asymptotic power law occurs at very long times. Thus, the maximum simulation time limits
how close to the dirty critical point one can still extract $z'$.

\subsubsection{Dirty critical point}

\begin{figure}
\includegraphics[width=\columnwidth]{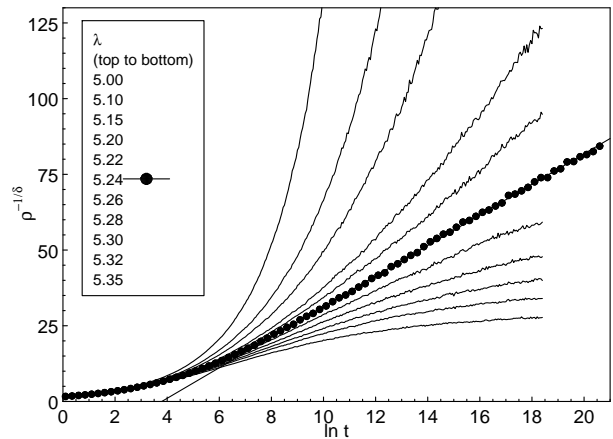}
\caption{$\rho^{-1/\bar\delta}$ vs. $\ln(t)$ for a system of $10^4$ sites with $p=0.3$
and $c=0.2$. The filled circles mark the critical birth rate $\lambda_c=5.24$, and the
straight line is a fit of the long-time behavior to eq.\ (\ref{eq:logdecay})}
\label{fig:criticalp03c02}
\end{figure}

After having discussed the Griffiths region, we now turn to the most interesting
parameter region, the vicinity of the dirty critical point. In contrast to the clean
critical birth rate $\lambda_c^0$ which is well known from the literature
\cite{Jensen93}, the dirty critical birth rate $\lambda_c$ is not known a priori. In
order to find $\lambda_c$ and at the same time test the predictions of the activated
scaling picture of section \ref{subsec:activated},  we employ the logarithmic time
dependence of the density, eq.\ (\ref{eq:logdecay}). In figure \ref{fig:criticalp03c02}
we plot $\rho^{-1/\bar\delta}$ with the predicted $\bar\delta=0.38197$ against $\ln(t)$
for a system of $10^4$ sites with $p=0.3$ and $c=0.2$.
Because the dynamics at the dirty critical point is expected to be extremely slow, we
have simulated up to $t=10^8$ ($10^9$ for the critical curve). As before, the data are
averages over 480 runs, each with a different disorder realization.

In this type of plot, the logarithmic time dependence (\ref{eq:logdecay}) is represented
by a straight line. Subcritical data should curve upward from the critical straight line,
while supercritical data should curve downward and eventually settle to a constant
long-time limit. From the data in figure \ref{fig:criticalp03c02} we conclude that the
dirty critical point indeed follows the activated scaling scenario associated with an
infinite-randomness critical point. The critical birthrate is $\lambda_c=5.24\pm 0.01$.
At this $\lambda$, the density follows eq.\ (\ref{eq:logdecay}) over almost four orders
of magnitude in $t$.

The statistical error of the plotted average densities can be estimated from the standard
deviation of $\rho(t)$ between the 480 separate runs. For the critical curve,
$\lambda=5.24$, in figure \ref{fig:criticalp03c02}, the error of the average density
remains below $0.002$ which corresponds to about a symbol size in the figure (at the
long-time end of the plot). We have also checked for possible finite-size effects by
repeating the calculation for a smaller system size of $L=10^3$. Within the statistical
error the results for $L=10^3$ and $L=10^4$ are identical, from which we conclude that
our data are not influenced by finite size effects.

\subsubsection{Universality}

We now turn to the questions of universality: Is the activated scaling
scenario valid for all impurity concentrations $p$ and strengths $c$, and is the value of
the critical exponent $\bar\delta$ the same for all cases? To answer these questions we
have repeated the above critical point analysis for different sets of the disorder
parameters $p$ and $c$. In this subsection, we first show that all these data are in
agreement with the activated scaling scenario with an universal exponent $\bar\delta$. We
then discuss whether they could interpreted in a conventional power-law scaling scenario
as well.

\begin{figure}
\includegraphics[width=\columnwidth]{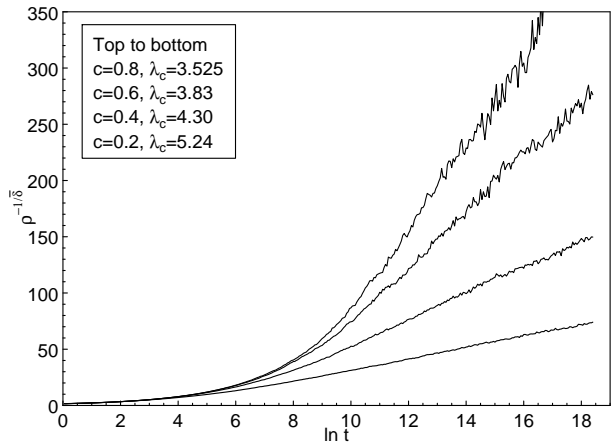}
\caption{Time evolution of the density for systems of $10^4$ sites with $p=0.3$ and
$c=0.2, 0.4, 0.6$ and 0.8 at their respective critical points plotted as
$\rho^{-1/\bar\delta}$ vs. $\ln(t)$ as in figure \ref{fig:criticalp03c02}.}
\label{fig:allcritical}
\end{figure}

In the first set of calculations we have kept the impurity concentration at $p=0.3$, but
we have varied their relative strength from $c=0.2$ to 0.8. In all cases, the density decay
at the respective dirty critical birth rate $\lambda_c$ follows the logarithmic law
(\ref{eq:logdecay}) with the predicted $\bar\delta=0.38197$
over several orders of magnitude in $t$. Figure \ref{fig:allcritical} shows these
critical curves for systems with $c=0.2, 0.4, 0.6$ and 0.8.
The seemingly larger fluctuations for the curves with higher $c$ are caused by the way
the data are plotted: The large negative exponent $-1/\bar\delta$ strongly stretches the
low-density part of the ordinate.

We have obtained analogous results from the second set of runs where we kept $c=0.2$
constant but varied the impurity concentration from $p=0.2$ to 0.5. From these simulation
results we conclude that the data are in agreement with the activated scaling scenario
for all studied parameter values including the case of weak disorder. (Note that for
$p=0.3, c=0.8$, the disorder-induced shift of the critical birthrate is small,
($\lambda_c -\lambda_c^0)/\lambda_c^0 \approx 0.07$.) Moreover, our results are
compatible with a universal value of 0.38197 for the exponent $\bar\delta$.

However, we would like to emphasize that while our data do \emph{not} show any indication
of non-universality, we cannot exclude some variation of $\bar\delta$ with the
disorder strength. This is caused by the fact that even though we observe the logarithmic
time dependence (\ref{eq:logdecay}) over almost four orders of magnitude in $t$, this
corresponds only to about a factor of 2 to 3 in $\ln(t)$. This is a very small range for
extracting the exponent of the power-law relation between $\rho$ and $\ln(t)$. More
specifically, the asymptotic critical time dependence of the density for $p=0.3, c=0.2$
can be fitted by $\rho = (A*\ln(t) + B)^{-0.38197}$ with $A\approx 5.12$ and $B\approx
-20.1$ (this is the straight line in figure \ref{fig:criticalp03c02}). Comparison with
eq.\ (\ref{eq:logdecay}) shows that $B$ represents a correction to scaling. For a
reliable extraction of the exponent one would want the $\ln(t)$ term to be at least one
order of magnitude larger than $B$ at the very minimum. This corresponds to $\ln(t)
\approx 40$ or $t \approx 10^{17}$ which is clearly unreachable in a Monte Carlo
simulation for the foreseeable future.

\begin{figure}
\includegraphics[width=\columnwidth]{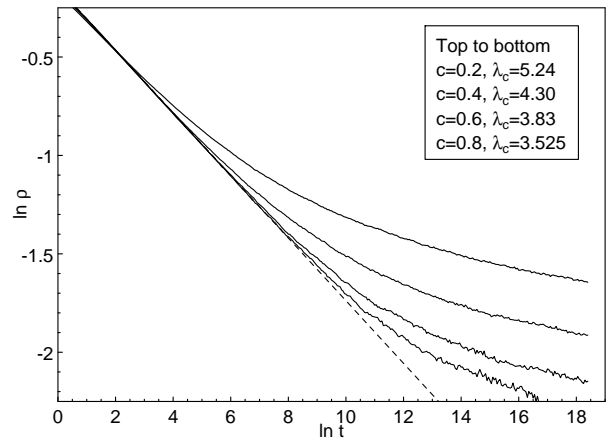}
\caption{Replot of the data of figure \ref{fig:allcritical} in log-log form. The dashed
line shows a power-law decay with the {\em clean} critical exponent $\delta=0.159$.}
\label{fig:allcritical_power}
\end{figure}

We now turn to the question of whether the numerical data could also be interpreted in
terms of conventional power-law scaling. To this end, we first
replot the critical density decay curves as identified above \cite{CPIDENT}
in standard log-log form in Fig.\
\ref{fig:allcritical_power}.
This figure shows that the density decay at early times follows a power law with the
\emph{clean} decay exponent $\delta=0.159$. However, after some disorder dependent
crossover time $t_x$ the curves show pronounced upward curvature. This curvature persists
to the largest times and signifies an asymptotic decay that is slower than any power law.
The crossover time $t_x$ which increases with decreasing disorder $c\to 1$ nicely agrees
with the onset of the asymptotic logarithmic behavior in Fig.\ \ref{fig:allcritical} for
all disorder strengths. Thus, the time dependence of the density $\rho(t)$ directly crosses
over from the clean critical (power law) behavior at short times to the activated
(logarithmic) behavior (\ref{eq:logdecay}) at large times. We find no indication of
any power laws other than the clean one, not even at transient times.
We thus conclude that our data, at least for
times up to $t=10^8$ are not compatible with power-law scaling at the dirty critical
point.

It would be highly desirable to discriminate between power-law and activated scaling by
performing a full scaling analysis of the data using relations
(\ref{eq:rho},\ref{eq:Ps},\ref{eq:N})
or (\ref{eq:rho_activated},\ref{eq:Ps_activated},\ref{eq:N_activated}), respectively.
Unfortunately, such an analysis is impossible because the extremely slow dynamics leads
to the abovementioned strong corrections to scaling in the accessible range of times
up to $t=10^9$.

\subsection{Time evolution starting from a single particle}

In addition to the simulations staring from a full lattice, we have also performed
simulations of the time evolution starting from a single active site in an otherwise
empty lattice. In these calculations, we have monitored the survival probability $P_s$
and the number of sites $N$ in the active cluster as functions of time. According to
eqs.\ (\ref{eq:logdecay}) and (\ref{eq:clustersize}), these quantities are expected to
behave as $P_s(t) \sim [\ln(t)]^{-1/\bar\delta}$ and $N(t) \sim [\ln(t)]^{\bar\Theta}$ at
the dirty critical point.

\begin{figure}
\includegraphics[width=\columnwidth]{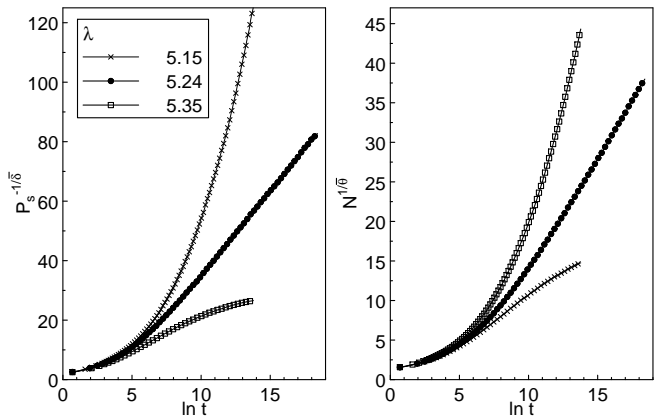}
\caption{Survival probability $P_s$ and particle number $N$ in the active cluster for a
simulation starting from single active site. The disorder parameters are $p=0.3$ and
$c=0.2$.} \label{fig:p03c02local}
\end{figure}

Figure \ref{fig:p03c02local} shows plots of $P_s^{-\bar\delta}$ and $N^{1/\bar\Theta}$
vs. $\ln(t)$ for a system with $p=0.3$ and $c=0.2$.
The data are averages over 480 different disorder realizations. For each realization we
have performed 200 runs starting from a single active site at a random position. The
system size $L=10^6$ was several orders of magnitude larger than the largest active
cluster, thus we have effectively simulated an infinite-size system. Figure
\ref{fig:p03c02local} shows that $P_s$ and $N$ indeed follow the predicted logarithmic
laws with $\bar\delta=0.38197$ and $\bar\Theta=1.236$ at the critical birthrate
$\lambda_c=5.24$. Subcritical and supercritical data curve away from the critical
straight lines as expected. Thus, the simulations starting from a single site also
confirm the activated scaling scenario resulting from an infinite-randomness critical
point.

\subsection{Steady state}

\begin{figure}
\includegraphics[width=\columnwidth]{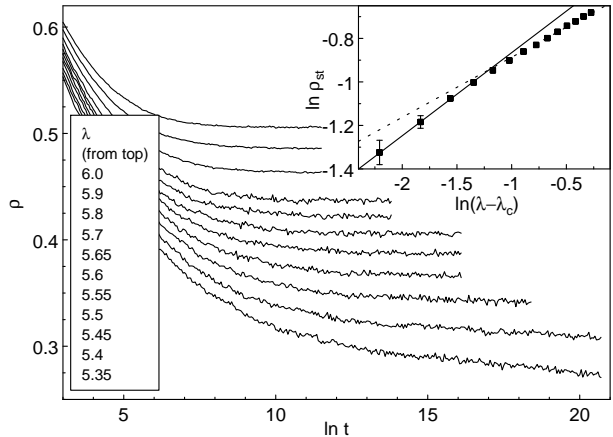}
\caption{Time evolution of the density in the active phase for a system with $p=0.3$ and
$c=0.2$. The system sizes are $10^4$ for $\lambda\ge 5.8$ and $10^3$ otherwise. Inset:
Steady state density vs. distance from critical point. The solid line is a fit of the
data for $\lambda-\lambda_c<0.3$ to the expected dirty power law $\rho_{\rm stat} \sim
(\lambda-\lambda_c)^\beta$ with $\beta=0.38197$. The dashed line is a fit of the data for
$\lambda-\lambda_c>0.3$ to a power law with the clean $\beta=0.2765$.}
\label{fig:p03c02steady}
\end{figure}

Lastly, we have studied the behavior of the steady state density in the active phase.
Close to the critical point it is expected to vary as $\rho_{\rm stat} \sim
(\lambda-\lambda_c)^\beta$ with $\beta=0.38197$. These calculations require a
particularly high numerical effort because the approach to the steady is logarithmically
slow close to the critical point. Therefore, we have simulated up to $t=10^9$ for
birthrates close to $\lambda_c$. Figure \ref{fig:p03c02steady} shows the time evolution
of the density in the active phase for a system with $p=0.3$ and $c=0.2$, averaged over
480 disorder realizations (one run per realization).
The steady state densities can be obtained by averaging the constant part of each of the
curves. The data clearly demonstrate the extremely slow approach to the steady state. For
$\lambda \le 5.4$ the steady state is not reached (within the statistical error) even
after $t = 10^9$. Therefore, our steady state density values for these $\lambda$ are only
rough estimates.

The inset of figure \ref{fig:p03c02steady} shows the resulting steady state densities
plotted vs. the distance from the critical point. This graph nicely demonstrates the
crossover from clean to dirty critical behavior. The data away from the critical point
($\lambda-\lambda_c>0.3$) can be well fitted by a power law with the clean value of the
critical exponent $\beta=0.2765$. When approaching the critical point the curve becomes
steeper, and for $\lambda-\lambda_c < 0.3$ the data can be reasonably well fitted by a
power law with the expected dirty critical exponent $\beta=0.38197$. The position of the
crossover can be compared to an estimate from the time dependence at $\lambda_c$ in
figure \ref{fig:criticalp03c02}. The critical curve reaches its asymptotic logarithmic
form at about $t_x \approx 10^4$ which corresponds to a crossover density of $\rho_x
\approx 0.3$ in rough agreement with the crossover density in the inset of figure
\ref{fig:p03c02steady}.

\section{Conclusions}
\label{sec:conclusions}

To summarize, we have presented the results of large-scale Monte-Carlo simulations of a
one-dimensional contact process with quenched spatial disorder for large systems of up to
$L=10^7$ sites and very long times up to $t=10^9$. These simulations show that the
critical behavior at the nonequilibrium phase transition is controlled by an
infinite-randomness fixed point with activated scaling and ultraslow dynamics, as
predicted in Ref.\ \cite{hooyberghs}. Moreover, the simulations provide evidence that
this behavior is universal. The logarithmically slow time dependencies
(\ref{eq:logdecay},\ref{eq:clustersize}) are valid (with the same exponent values) for
all parameter sets investigated including the weak-disorder case. However, the approach
to this universal asymptotic behavior is extremely slow, with crossover times of the
order of $10^4$ or larger. Since most of the earlier Monte-Carlo simulations did not
exceed $t=10^5 \ldots 10^6$ this may explain why nonuniversal (effective) exponents were
seen in previous work. We have also presented results for the Griffiths region between
the clean and the dirty critical points. Here, we have found power-law dynamical behavior
with continuously varying exponents, in agreement with theoretical predictions
\cite{Noest86,Noest88}.

We now discuss the relation of our results to a more general theory of rare region
effects at phase transitions with quenched disorder. In Ref.\ \cite{VojtaSchmalian05} a
general classification of phase transitions in quenched disordered systems with
short-range interactions \cite{RKKY} has been suggested, based on the effective
dimensionality $d_{\rm eff}$ of the rare regions. Three cases can be distinguished.

(i) If $d_{\rm eff}$ is below the lower critical dimension $d_c^-$ of the problem, the
rare region effects are exponentially small because the probability of a rare region
decreases exponentially with its volume but the contribution of each region to
observables increases only as a power law. In this case, the critical point is of
conventional power-law type. Examples in this class include, e.g., the classical
equilibrium Ising transition with point defects where $d_{\rm eff}=0$ and $d_c^-=1$.

(ii) In the second class, with $d_{\rm eff}=d_c^-$, the Griffiths effects are of
power-law type because the exponentially rarity of the rare regions in $L_r$ is overcome
by an exponential increase of each region's contribution. In this class, the critical
point is controlled by an infinite-randomness fixed point with activated scaling.
Examples include the quantum phase transition in the random transverse field Ising model
($d_{\rm eff}=d_c^-=1)$ \cite{dsf9295,taufootnote} as well as the transition discussed in
this paper (where $d_{\rm eff}=d_c^-=0$).

(iii) Finally, for  $d_{\rm eff}>d_c^-$, the rare regions can undergo the phase
transition independently from the bulk system. This leads to a destruction of the sharp
phase transition by smearing. This behavior occurs at the equilibrium Ising transition
with plane defects where $d_{\rm eff}=2,d_c^-=1$ \cite{us_planar}, the quantum phase
transition in itinerant magnets \cite{us_rounding}, and for the contact process with
extended (line or plane) defects \cite{Dickison05,contact_pre}.

Thus, the results of this paper do fit into the general rare-region based classification
scheme of phase transitions with quenched disorder and short-range interactions
\cite{VojtaSchmalian05}. These arguments also suggest that the behavior of the phase
transition in a higher-dimensional disordered contact process should be controlled by an
infinite-randomness fixed point as well.

We conclude by pointing out that the unconventional behavior found in this paper may
explain the striking absence of directed percolation scaling \cite{hinrichsen_exp} in at
least some of the experiments. However, the extremely slow dynamics will prove to be a
challenge for the verification of the activated scaling scenario not just in simulations
but also in experiments.

\section*{Acknowledgements}
This work has been supported in part by the NSF under grant nos. DMR-0339147 and
PHY99-07949 as well as by the University of Missouri Research Board. Thomas Vojta is a
Cottrell Scholar of Research Corporation. We are also grateful for the hospitality of the
Aspen Center for Physics and the Kavli Institute for Theoretical Physics, Santa Barbara
during the early stages of this work.


\begin{thebibliography}{99}
\bibitem{betaprime} At more general absorbing state transitions, e.g., with several
    absorbing states, the survival probability scales with an exponent $\beta'$ which may
    be different from $\beta$ (see, e.g., \cite{Hinrichsen00}).
\bibitem{hyperscaling} This relation relies on hyperscaling; it is only valid below the
    upper critical dimension $d_c^+$, which is four for directed percolation.
\bibitem{RKKY} In the case of long-range interactions, rare region effects can be enhanced
     further by the interactions between the rare regions. This happens, e.g., in
     the important case of magnetic transitions in metals due to the RKKY interaction
     \cite{vlad05}.
\bibitem{taufootnote} At a zero temperature quantum phase transition, imaginary time acts
     as an extra dimension. Because the quenched disorder is time-independent, the rare
     regions are effectively one-dimensional.
\end{thebibliography}
\end{document}